\documentclass[twocolumn,preprintnumbers,amsmath,amssymb,prb]{revtex4}
\usepackage{graphicx}% Include figure files
\usepackage{dcolumn}% Align table columns on decimal point
\usepackage{bm}% bold math

\usepackage[usenames,dvipsnames]{color}
\definecolor{blue}{rgb}{0.,.0,1.0} 

\def \be {\begin{equation}}
\def \ee {\end{equation}}
\def \ben {\begin{eqnarray}}
\def \een {\end{eqnarray}}
\newcommand \comma {\mbox{\makebox[.1 in]{ },}}

\newcommand \perd {\mbox{\makebox[.1 in]{ }.}}

\newcommand \dpt {{\mathcal D}}
\def \be {\begin{equation}}
\def \ee {\end{equation}}
\def \ben {\begin{eqnarray}}
\def \een {\end{eqnarray}}

\begin{document}
\draft

\title{Generalized quantum Fokker-Planck equation for photoinduced nonequilibrium  processes with positive definiteness condition}

\author{Seogjoo Jang\footnote{Email:sjang@qc.cuny.edu}}
\affiliation{Department of Chemistry and Biochemistry, Queens College, City University of New York, 65-30 Kissena Boulevard, Queens, NY 11367\footnote{mailing address} \& PhD programs in Chemistry and Physics, and Initiative for Theoretical Sciences, Graduate Center, City University of New York, 365 Fifth Avenue, New York, NY 10016}

\date{Revised and  resubmitted to the Journal of Chemical Physics on January 7, 2016 }

\begin{abstract}
This work provides a detailed derivation of a generalized quantum Fokker-Planck equation (GQFPE) appropriate for photo-induced quantum dynamical processes. The path integral method pioneered by Caldeira and Leggett (CL) [Caldeira and Leggett, Physica A {\bf 121}, 587 (1983)] is extended for  a nonequilibrium influence functional, which has been obtained for general cases where the ground and the excited electronic state baths can be different.  Both nonequilibrium and non-Markovian effects are accounted for consistently by expanding the paths in the exponents of the influence functional with respect to time up to the second order.  This procedure results in approximations involving only single time integrations for the exponents of the influence functional but with additional time dependent boundary terms that have been ignored in previous works.  The boundary terms complicate the derivation of a time evolution equation, but do not affect position dependent physical observables or the dynamics in the steady state limit.  For an effective density operator with the boundary terms factored out, a time evolution equation is derived through short time expansion of the effective action followed by Gaussian integrations in analytically continued complex domain of space.  This leads to a compact form of GQFPE with time dependent kernels and additional terms, which make the resulting equation the Dekker form [H. Dekker, Phys. Rep. {\bf 80}, 1 (1981)].  Major terms of the equation are analyzed for the case of Ohmic spectral density with Drude cutoff, which shows that the new GQFPE satisfies the positive definiteness condition in medium to high temperature limit.
\end{abstract}

\maketitle

\vskip 0.5in

%%\section{Outline}
%%\tableofcontents

%\vskip 0.5in

%\narrowtext
\section{Introduction}
The concept of Brownian motion,\cite{gardiner,vankampen} or more specifically, Langevin equation,\cite{langevin-cr146} was originally developed  under the premise that the system of interest follows fully deterministic paths if left alone and that environmental effects can be accounted for by random forces and frictional drags satisfying the fluctuation-dissipation relationship.   How to extend such description to the quantum mechanical regime  governing time evolution of quantum operators had remained difficult in practice\cite{hanggi-chaos15,grabert-cp322} or was even considered impossible,\cite{vankampen-jpcb109} although some advances have been made.\cite{hanggi-chaos15,tsusaka-pre59,kleinert-pla200,datta-jpca109}  On the other hand, the Fokker-Planck equation (FPE),\cite{risken-fpe} which considers time evolution of distribution function instead, is more amenable to quantum generalization because the distribution can be obtained naturally from a quantum density operator retaining complete information on the quantum system.\footnote{The definition of QFPE seems to be somewhat subjective and can differ for each researcher.  Here, we define it loosely as a time evolution equation for quantum particle approaching the FPE in the classical limit.  Thus, it can be viewed as a kind of quantum master equation for a continuous quantum degree of freedom.}    Indeed, a few well defined and tractable derivations of quantum FPE (QFPE)\cite{caldeira-pa121,diosi-pa199,kuhn-jcp112,vacchini-prl84,vacchini-pre66,yano-jsp158,banik-pre65,grabert-cp322} or hierarchical QFPEs\cite{tanimura-pre47,tanimura-jcp107,yan-arpc56,tanimura-jcp142} are available now.  

One of the most well-known derivation of QFPE was provided by Caldeira and Leggett (CL)\cite{caldeira-pa121} based on the Feynman-Vernon influence functional formalism.\cite{feynman-qmech,weiss}  This approach has also been extended to the case of nonadiabatic quantum dynamics by Garg, Onuchic, and Ambegaokar (GOA).\cite{garg-jcp83}  CL's derivation of QFPE invokes high temperature and the Markovian approximation for the bath dynamics.  Although not explicit, an assumption of weak system-bath coupling appears to be implicit in the derivation as well.  Indeed, a distinct quantum Smoluchowski equation (QSE)\cite{pechukas-ap9,ankerhold-prl87,grabert-cp322} is obtained following a similar approach but taking the effect of strong system-bath coupling properly.  However, this QSE still assumes that the bath relaxes much faster than the system, and does not account for the non-Markovian effect that can have potentially important effects. 

Major applications of QFPE include quantum extension\cite{palchikov-pa316,banerjee-pre65,banik-pre65} of Kramers' barrier crossing problem and proton or electron transfer dynamics.\cite{garg-jcp83,yang-jsp74,basilevsky-jcsft93,casado-pascual-jcp118,jung-jpca103}   In particular, for the latter case, there has been growing interest in the study of fast photo-induced reaction dynamics that can occur during time scales comparable to those of molecular relaxation and dephasing dynamics.\cite{adv-et,cho-jcp103,wang-science316} For these, currently available QFPE or QSE are not well suited.  Thus, generalization of QFPE to include non-Markovian and nonequilibrium effects remains an important and interesting theoretical issue to be addressed.   Although more general hierarchical equations\cite{tanimura-pre47,tanimura-jcp107,yan-arpc56,tanimura-jcp142} may be used to this end, the benefit of having a single closed form equation, which  can account for such non-Markovian and nonequilibrium effects, cannot be overestated.  

In addition, the fact that CL's QFPE\cite{caldeira-pa121} is not positive definite remains a lingering theoretical issue.  Although it is true that positive definiteness is not necessary for accurate description of the open system quantum dynamics,\cite{pechukas-prl73,grabert-cp322} it is still important to understand the source of its violation and how to fix the problem.  Di\'{o}si showed that the non-positivity can result from an inconsistent application of the Markovian approximation\cite{diosi-pa199} and that it can partially be corrected by including next order terms in the intermediate temperature regime.    The present work shows that a a similar consideration can be made in deriving a generalized QFPE (GQFPE) and that the resulting equation is of the Dekker form,\cite{diosi-pa199,dekker-pr80} which has a well-defined condition for positive definiteness.  A detailed consideration of this equation for Ohmic spectral density with Drude cutoff shows that the Lindblad's positive definiteness condition\cite{lindblad-cmp48} can indeed be satisfied in the steady state limit under reasonable physical condition.

The paper is organized as follows.  Section II presents the main theoretical development based on the standard path integral formulation.  Section III provides numerical analysis of major terms of newly derived GQFPE for the case of Ohmic spectral density with Drude cutoff. Section IV concludes the paper by summarizing the main results and their implication.

\section{Theory}
Consider the following total Hamiltonian: 
\be
\hat H=\hat H_g|g\rangle\langle g|+\hat H_e|e\rangle\langle e|\comma
\ee
where $|g\rangle$ is the ground electronic state and $|e\rangle$ is the excited electronic state. $\hat H_g$ and $\hat H_e$ are nuclear Hamiltonians in respective ground and electronic states, and have the following forms:
\ben
&&\hat H_g=\frac{\hat p^2}{2m}+V_g(\hat q)\nonumber \\
&&\hspace{.1in}+\sum_\alpha \left \{\frac{\hat p_\alpha^2}{2m_\alpha}+\frac{m_\alpha \omega_\alpha^2}{2}\left (\hat x_\alpha-\frac{c_{\alpha,g}}{m_\alpha \omega_\alpha^2}\hat q\right )^2\right\} \label{eq:hg} \comma\\
&&\hat H_e=\frac{\hat p^2}{2m}+V_e(\hat q)\nonumber \\
&&\hspace{.1in}+\sum_\alpha \left \{\frac{\hat p_\alpha^2}{2m_\alpha}+\frac{m_\alpha \omega_\alpha^2}{2}\left (\hat x_\alpha-\frac{c_{\alpha,e}}{m_\alpha \omega_\alpha^2}\hat q\right )^2\right\}\label{eq:hg} \perd
\een
In the above expression, $\hat q$ and $\hat p$ represent the position and momentum operators of the quantum nuclear degree of freedom of the system,  and $\hat x_\alpha$'s and $\hat p_\alpha$'s represent the position and momentum operators of all the bath modes bilinearly coupled to the system.    The system nuclear degree of freedom is assumed to be one dimensional here, but extension of the present work for multidimensional situation is straightforward. 

It is assumed that there is no coupling between the ground and the excited state in the absence of radiation.  
The total density operator at time $t$ is denoted as $\hat \rho_T(t)$. As the initial condition at $t=0$, we consider the situation where the entire system plus bath degrees of freedom are prepared at their canonical equilibrium for the ground electronic state as follows: 
\be
\hat \rho_T(0)=|g\rangle\langle g|\hat \rho_g=|g\rangle\langle g|e^{-\beta \hat H_g}/Tr\{e^{-\beta \hat H_g}\}  \ .
\ee

Given that an impulsive excitation is applied to the system at time zero and under the Condon approximation that the transition dipole is independent of nuclear coordinates, a vertical transition from $|g\rangle$ 
to $|e\rangle$ occurs while all other degrees of freedom remain frozen. Thus, the total density operator for $t\geq 0+$, following an impulsive excitation at $t=0$, 
is given by 
\be
\hat \rho_T(t)=\hat \rho_e(t)|e\rangle\langle e|=e^{-i\hat H_e t/\hbar}\hat \rho_g e^{i\hat H_e t/\hbar}|e\rangle\langle e| \, ,
\ee
where $\hat \rho_e (t)$ is the total density operator representing the system nuclear coordinate in the excited electronic state and the bath.  
Taking the trace of this over the bath degrees of freedom, we obtain the reduced density operator describing the system nuclear degree of 
freedom as follows: 
\be
\hat \sigma_e(t)=Tr_b\{\hat \rho_e(t)\}=Tr_b\{e^{-i\hat H_e t/\hbar}\hat \rho_g e^{i\hat H_e t/\hbar}\} \comma \label{eq:sigmae-def}
\ee
where $Tr_b$ represents trace over the bath.  
For the derivation of GQFPE governing time evolution of $\hat \sigma_e (t)$, we extend the path integral approach developed by CL,\cite{caldeira-pa121} which has also been adopted by GOA\cite{garg-jcp83} for nonadiabatic quantum dynamical processes.

\subsection{Path integral representation and short time expansion}
The path integral representation for the reduced system density operator in the excited electronic state, $\hat \sigma_e(t)$ defined by Eq. (\ref{eq:sigmae-def}), can be found by utilizing standard expressions for both the real and imaginary time propagators.  The major steps are described in Appendix A, where the final expression for $\hat \sigma_e(t)$ is given by Eq. (\ref{eq:pt2}).  This expression can be simplified by introducing the following three spectral densities of the bath: 
\ben
&&\eta_e(\omega)\equiv\frac{\pi}{2}\sum_\alpha \frac{c_{\alpha,e}^2}{m_\alpha \omega_\alpha}\delta (\omega-\omega_\alpha)\comma \label{eq:eta_e}\\
&&\eta_c(\omega)\equiv\frac{\pi}{2}\sum_\alpha \frac{c_{\alpha,e}c_{\alpha,g}}{m_\alpha \omega_\alpha}\delta (\omega-\omega_\alpha)\comma\\
&&\eta_g(\omega)\equiv\frac{\pi}{2}\sum_\alpha \frac{c_{\alpha,g}^2}{m_\alpha \omega_\alpha}\delta (\omega-\omega_\alpha)\ .
\een
Then, with the definitions of Eqs. (\ref{eq:w1t})-(\ref{eq:w4t}), we can introduce the following nonequilibrium influence functional:
\ben
&&J[q'(\cdot),q''(\cdot),q_g(\cdot); t,\beta\hbar] =Z_b\ \exp\Big\{\frac{i}{\hbar} W_{e,I}(t) \nonumber \\ 
&&\hspace{.5in}-\frac{1}{\hbar}W_{e,R}(t) +\frac{i}{\hbar} W_c(t)+\frac{1}{\hbar}W_g(\beta\hbar \} \Big\} \ , \label{eq:inf-def}
\een
where $Z_b=\prod_\alpha \big(2\sinh(\omega_\alpha\beta\hbar/2)\big)^{-1}$,  $q'(\cdot)$ and $q''(\cdot)$ represent real time paths in the excited electronic state, and $q_g(\cdot)$ the imaginary time path in the ground electronic state.  Different exponents in the above nonequilibrium influence functional represent effective actions coming from different sources of  the system-bath interaction.  $W_{e,I}(t)$ and $W_{e,R}(t)$ are imaginary and real components of the contribution from the bath dynamics in the excited state, $W_c(t)$ represents coupling between the two baths in the excited and ground electronic states, and $W_g(\beta\hbar)$ is the imaginary time action due to thermal distribution of the bath in the ground electronic state.

With the definition of Eq. (\ref{eq:inf-def}), Eq. (\ref{eq:pt2}) can now be expressed as
\ben
&&\hat \sigma_e(t)=\frac{1}{Z_g} \int dq_i'  \int dq_i''   \int_{q_i'}^{q_i''}\dpt q_g(\cdot)\int dq_f'\int dq_f''\ \nonumber \\
&&\hspace{.1in}\times \int_{q_i'}^{q_f'} {\mathcal D} q'(\cdot)\int_{q_i''}^{q_f''} {\mathcal D}q''(\cdot) J[q'(\cdot),q''(\cdot),q_g(\cdot); t,\beta\hbar]   \nonumber \\
&&\hspace{.2in}\times e^{\frac{i}{\hbar}S_e[q'(\cdot);t]-\frac{i}{\hbar}S_e[q''(\cdot);t]-\frac{1}{\hbar}S_g^E[q_g(\cdot);\beta\hbar]} |q_f'\rangle\langle q_f''| \ .   \label{eq:sigmae-1}
\een
The nonequilibrium influence functional appearing in the above equation, as defined by Eq. (\ref{eq:inf-def}), is a direct extension of the Feynman and Vernon influence functional\cite{feynman-qmech,weiss} to the general case where the Hamiltonian for the initial equilibrium distribution can be different from that of the dynamics. Numerical evaluation of this is feasible extending novel computational methods,\cite{makri-arpc50} which is not the main focus here. 

For the derivation of GQFPE, let us  introduce the following time dependent kernels:
\ben
&&\tilde \eta_{e,I}(t)\equiv\frac{1}{\pi}\int_0^\infty d\omega\ \eta_e(\omega)\sin(\omega t) \ , \label{eq:eta-it}\\
&&\tilde \eta_{e,R}(t)\equiv\frac{1}{\pi}\int_0^\infty d\omega\ \eta_e(\omega)\coth(\frac{\omega \beta \hbar}{2})\cos(\omega t)\ , \label{eq:eta-rt}\\
&&\tilde \eta_c (t;q_g(\cdot))\equiv \frac{1}{\pi}\int_0^{\beta\hbar} d\tau \int_0^\infty d\omega\ \eta_c(\omega)\  \nonumber \\ 
&&\hspace{.7in}\times \frac{\cosh(\omega(\tau-it-\beta\hbar/2))}{\sinh(\omega\beta\hbar/2)}q_g(\tau)\perd
\een
Then, the three time dependent exponents in Eq. (\ref{eq:inf-def}), which are defined by Eqs. (\ref{eq:w1t})-(\ref{eq:w3t}), 
can be expressed as
\ben
&&W_{e,I}(t)=\int_0^t dt_1 \int_0^{t_1} dt_2\ \tilde \eta_{e,I}(t_2) (q'(t_1)-q''(t_1)) \nonumber \\
&&\hspace{1in}\times    (q'(t_1-t_2)+q''(t_1-t_2)) \ ,  \label{eq:w1t-1}\\
&&W_{e,R}(t)=\int_0^t dt_1\int_0^{t_1} dt_2\ \tilde \eta_{e,R}(t_2)(q'(t_1)-q''(t_1))\nonumber \\
&&\hspace{1in}\times (q'(t_1-t_2)-q''(t_1-t_2))\ ,  \label{eq:w2t-1}\\
&&W_c(t;q_g(\cdot))=\int_0^tdt_1\ \tilde\eta_c(t;q_g(\cdot)) (q'(t_1)-q''(t_1))\ .   \label{eq:w3t-1}
\een
Equations (\ref{eq:w1t-1}) and (\ref{eq:w2t-1}) above involve double time integrations, which need to be converted to single time integrations\cite{caldeira-pa121,garg-jcp83} for the derivation of GQFPE.   

Under the assumption that the decay of $\tilde \eta_{e,I}(t)$ is fast enough, $q'(t_1-t_2)$ and $q''(t_1-t_2)$ in Eq. (\ref{eq:w1t-1}) can be approximated with their second order expansions with respect to $t_2$ around $t_1$. The resulting expression can then be converted to single time integration through partial integration.  Thus, Eq. (\ref{eq:w1t-1}) can be approximated as
\ben
&&W_{e,I}(t)\approx  \int_0^t dt_1 {\mathcal K}_I^{(0)}(t_1)(q'(t_1)^2-q''(t_1)^2)\nonumber \\
&&-\int_0^t dt_1 \left ({\mathcal K}_I^{(1)}(t_1) +\frac{1}{2}\dot {\mathcal K}_I^{(2)}(t)\right)\nonumber \\
&&\hspace{.5in}\times (q'(t_1)-q''(t_1))(\dot{q}'(t_1)+\dot {q}''(t_1))  \nonumber  \\ 
&&-\frac{1}{2}\int_0^t dt_1 {\mathcal K}_I^{(2)} (t_1) (\dot q'(t_1)^2-\dot q''(t_1)^2) \nonumber \\ 
&&\left .+\frac{1}{2}{\mathcal K}_I^{(2)}(t_1) (q'(t_1)-q''(t_1))(\dot q'(t_1)+\dot q''(t_1))\right |_0^t \ , \label{eq:w1t-2}
\een
where the single dot over $q'(t)$ and $q''(t)$ denotes the first derivatives with respect to time and 
\be
{\mathcal K}_I^{(n)}(t_1)\equiv \int_0^{t_1} dt_2 \tilde \eta_{e,I}(t_2)t_2^n \perd \label{eq:kint}
\ee

Similarly, Eq. (\ref{eq:w2t-1}) can be approximated as
\ben
&&W_{e,R}(t)\approx\int_0^t dt_1 {\mathcal K}_R^{(0)}(t_1)(q'(t_1)-q''(t_1))^2 \nonumber \\
&&-\int_0^t dt_1 \left ({\mathcal K}_R^{(1)}(t_1)+\frac{1}{2} \dot {\mathcal K}_R^{(2)}(t_1)\right ) \nonumber \\
&&\hspace{.5in}\times (q'(t_1)-q''(t_1))(\dot{q}'(t_1)-\dot {q}''(t_1))\nonumber \\
&&-\frac{1}{2}\int_0^t dt_1 {\mathcal K}_R^{(2)}(t_1) (\dot q'(t_1)-\dot q''(t_1))^2\nonumber \\
&&\left .+\frac{1}{2}{\mathcal K}_R^{(2)}(t_1) (q'(t_1)-q''(t_1))(\dot q'(t_1)-\dot q''(t_1))\right |_0^t \ , \label{eq:w2t-2}
\een
where
\be
{\mathcal K}_R^{(n)}(t_1)\equiv \int_0^{t_1} dt_2 \tilde \eta_{e,R}(t_2)t_2^n \perd \label{eq:krnt}
\ee

With Eqs. (\ref{eq:w1t-2}) and (\ref{eq:w2t-2}), Eq. (\ref{eq:inf-def}) can be converted to an expression that involves only single time integrations, but with additional boundary terms.  Before presenting the final form, let us first collect all the contributions to Eq. (\ref{eq:inf-def}) from the boundary terms in Eqs. (\ref{eq:w1t-2}) and (\ref{eq:w2t-2}), and define
\ben
&&{\mathcal A}(r', r'',\dot r',\dot r'',t)=\nonumber \\
&&\hspace{.5in}\exp \Big \{-\frac{1}{2\hbar} {\mathcal K}_R^{(2)}(t) (q'-q'')(\dot q'-\dot q'') \nonumber \\
&&\hspace{.7in}+\frac{i}{2\hbar}{\mathcal K}_I^{(2)}(t) (q'-q'')(\dot q' +\dot q'')\Big\} \ . \label{eq:boundary-term}
\een
Collecting all the contributions from single time integration terms in Eqs. (\ref{eq:w1t-2}) and (\ref{eq:w2t-2}), let us also define
\ben
&&J_{eff}[q'(\cdot),q''(\cdot),q_g(\cdot);t,t_0]=\nonumber \\
&&\hspace{1in}\exp\left\{-\frac{1}{\hbar}{\mathcal C}_R[q'(\cdot),q''(\cdot);t,t_0]\right .\nonumber \\
&&\hspace{1.1in}\left .-\frac{i}{\hbar} {\mathcal C}_I[q'(\cdot),q''(\cdot);t,t_0]\right\}\ , \label{eq:jeff-def}
\een
where
\ben
&&{\mathcal C}_R[q'(\cdot),q''(\cdot);t,t_0]=\int_{t_0}^t dt_1\Big\{ {\mathcal K}_R^{(0)}(t_1)(q'(t_1)-q''(t_1))^2  \nonumber \\
&&\hspace{.1in} -\tilde {\mathcal K}_R^{(1)}(t_1)(q'(t_1)-q''(t_1))(\dot{q}'(t_1)-\dot{q}''(t_1)) \nonumber \\ 
&&\hspace{.1in}-\frac{1}{2}{\mathcal K}_R^{(2)}(t_1) (\dot q'(t_1)-\dot q''(t_1))^2\Big \} \ ,\\
&&{\mathcal C}_I[q'(\cdot),q''(\cdot);t,t_0]= \nonumber \\
&&\hspace{.1in}\int_{t_0}^t dt_1 \tilde {\mathcal K}_I^{(1)}(t_1)(q'(t_1)-q''(t_1))(\dot{q}'(t_1)+\dot{q}''(t_1)) \ ,  
\een
with
\ben
&&\tilde {\mathcal K}_R^{(1)}(t_1)={\mathcal K}_R^{(1)}(t_1)+\frac{1}{2}\dot {\mathcal K}_R^{(2)}(t_1) \ , \label{eq:krtil1}\\
&&\tilde {\mathcal K}_I^{(1)}(t_1)={\mathcal K}_I^{(1)}(t_1)+\frac{1}{2}\dot {\mathcal K}_I^{(2)}(t_1) \ . \label{eq:kitil1}
\een
Finally, the following effective time dependent action can be introduced:
\ben
&&S_{eff}[q(\cdot)q_g(\cdot);t,t_0]\nonumber \\
&&=\int_{t_0}^t dt_1 \Big\{ \frac{m_{e}(t)}{2}\dot{q}(t_1)^2 -U_e(q(t_1),q_g(\cdot);t_1) \Big\}\ ,
\een
where
\ben
&&m_{e}(t)=m -{\mathcal K}_I^{(2)}(t)\ ,  \label{eq:me_def}\\
&&U_e(q, q_g(\cdot),t)=V_e(q)+\left (\frac{\kappa_{e}}{2}-{\mathcal K}_I^{(0)}(t)\right) q^2 \nonumber \\
&&\hspace{.8in}  - \tilde \eta_c(t,q_g(\cdot)) q \ . \label{eq:ue_def}
\een
In the above expression, $\kappa_{e}$, which is defined by Eq. (\ref{eq:lam_e}), is an effective harmonic oscillator spring constant due to the bath in the excited electronic state.  
Then, the position space matrix element of the reduced density operator, Eq. (\ref{eq:sigmae-1}), can be expressed as follows:
\ben
&&\langle q_f'|\hat \sigma_e(t)|q_f''\rangle = \int dq_i'\int dq_i'' \int_{q_i'}^{q_i''} \dpt q_g(\cdot)  {\mathcal P}[q_g(\cdot);\beta\hbar] \nonumber \\
&&\times \int_{q_i'}^{q_f'}\dpt q'(\cdot)\int_{q_i''}^{q_f''}\dpt q''(\cdot) {\mathcal A}(q_f',q_f'',\dot q_f',\dot q_f'',t)\nonumber \\
&&\hspace{.3in}\times  J_{eff}[q'(\cdot),q''(\cdot);t,0]\nonumber \\
&&\hspace{.3in}\times  e^{\frac{i}{\hbar}S_{eff}[q'(\cdot),q_g(\cdot);t,0]-\frac{i}{\hbar}S_{eff}[q''(\cdot),q_g(\cdot);t,0]}\ , \label{eq:sigmae-0}
\een
where the fact that ${\mathcal A}(q',q'',\dot q',\dot q'',0)=1$ has been used and ${\mathcal P}[q_g (\cdot);\beta\hbar]$ is the probability density for the imaginary time path with the following expression:
\ben
&&{\mathcal P}[q_g(\cdot);\beta\hbar]=\frac{Z_b}{Z_g}\exp\left\{-\frac{1}{\hbar}S_g^E[q_g(\cdot);\beta\hbar] \right . \nonumber \\
&& +\frac{1}{\hbar}\int_0^{\beta \hbar}d\tau\int_0^{\tau}d\tau_1 \int_0^{\infty}\frac{d\omega}{\pi} \eta_g(\omega) \nonumber \\
&&\hspace{.2in}\left . \times \frac{\cosh(\omega(\tau-\tau_1-\beta\hbar/2))}{\sinh (\omega\beta\hbar/2)}q_g(\tau_1)q_g(\tau)\right\}\ . 
\een

Equation (\ref{eq:sigmae-0}) is the best form available for deriving a  GQFPE.  Note the presence of the time dependent prefactor ${\mathcal A}(q_f',q_f'',\dot q_f',\dot q_f'',t)$, which comes from the boundary values of time integrations.  This term vanishes for $q_f'=q_f''$ at all time or in the long time limit where ${\mathcal K}_R^{(2)}(t)$ and ${\mathcal K}_I^{(2)}(t)$ decay to zero.  Thus, it does not contribute to the calculation of position dependent observables at any time or any observables in the steady state limit where the initial memory of the bath disappears.  However, for general situations, it remains as a source of ambiguity in deriving the time evolution equation and has not been considered  in previous treatments by CL,\cite{caldeira-pa121} GOA,\cite{garg-jcp83} and Di\'{o}si\cite{diosi-pa199} who all considered only the Markovian or steady state limit.    

\subsection{Time evolution equation}
Let us define the time dependent part in Eq. (\ref{eq:sigmae-0}) except for the prefactor and the ground state influence functional as follows:
\ben
&&\tilde\sigma_e(q_f',q_f'';t) \equiv \int_{q_i'}^{q_f'}\dpt q'(\cdot)\int_{q_i''}^{q_f''}\dpt q''(\cdot) \nonumber\\
&&\hspace{.3in}\times  J_{eff}[q'(\cdot),q''(\cdot);t,0] \nonumber \\
&&\hspace{.3in}\times  e^{\frac{i}{\hbar}S_{eff}[q'(\cdot),q_g(\cdot);t,0]-\frac{i}{\hbar}S_{eff}[q''(\cdot),q_g(\cdot);t,0]}\ .  \label{eq:rhoe}
\een
In the above expression, dependences of $\tilde \sigma_e$ on $q'$, $q''$, and $q_g(\cdot)$ have not been shown explicitly.
A time evolution equation for $\tilde \sigma_e(q_f',q_f'';t)$ can be derived employing the short time expansion of path integral expression as was done by CL\cite{caldeira-pa121} and GOA.\cite{garg-jcp83}  

Due to the fact that $J_{eff}[q'(\cdot),q''(\cdot);t,0]$ defined by Eq. (\ref{eq:jeff-def}) now involves single time integration in the exponent, it can be expressed as the product of discretized terms as follows:
\ben
J_{eff}[q'(\cdot),q''(\cdot);t+\delta t,0]&=&J_{eff}[q'_\delta(\cdot),q''_\delta(\cdot);t+\delta t,t] \nonumber \\
&&\times J_{eff}[q'_\delta(\cdot),q''_\delta(\cdot);t,t-\delta t] \nonumber\\ 
&&\hspace{.5in}\cdots    \nonumber \\
&&\times J_{eff}[q'_\delta(\cdot),q''_\delta(\cdot);\delta t,0]  \ .\nonumber \\
\een
Similar expressions can be found for $e^{\frac{i}{\hbar}S_{eff}[q'(\cdot),q_g(\cdot);t,0]}$ and $e^{-\frac{i}{\hbar}S_{eff}[q''(\cdot),q_g(\cdot);t,0]}$ as well.  Therefore, approximating the paths $q'(\cdot)$ and $q''(\cdot)$ by a collection of  discretized paths $q'_\delta(\cdot)$'s and $q''_\delta (\cdot)$'s with time interval $\delta t$, and assuming that $m_e(t)$ remains virtually constant during each time interval $\delta t$, we can express Eq. (\ref{eq:rhoe}) as follows:
\ben 
&&\tilde \sigma_e(q_f',q_f'';t+\delta t) = \frac{m_e}{2\pi \hbar t}\int dq'\int dq'' \tilde \sigma_e(q',q'';t)\nonumber \\ 
&&\hspace{.2in}\times J_{eff}[q'(\cdot),q''(\cdot);t+\delta t,t] \nonumber \\
&&\hspace{.2in}\times  e^{\frac{i}{\hbar}S_{eff}[q'(\cdot);t+\delta t, t]-\frac{i}{\hbar}S_{eff}[q''(\cdot);t+\delta t,t]} \ , \label{eq:sigmae-path}
\een
where the standard normalization factor of $\sqrt{m_e/(2\pi \hbar t)}$ was used for the path integral.  
Let us introduce $\delta q'=q_f'-q'$, $\delta q''=q_f''-q''$, $\bar q'=(q_f'+q')/2$, and $\bar q''=(q_f''+q'')/2$.  Then, assuming that ${\mathcal K}_I^{(n)}(t)$ and ${\mathcal K}_R^{(n)}(t)$ also remain virtually constant during the time interval of $\delta t$ and approximating the trajectories as straight lines, we obtain the following expressions:
\ben
&&S_{eff}[q'(\cdot);t+\delta t, t]\approx \frac{m_e(t)}{2\delta t}\delta q'^2-\delta t U_e(\bar q',t) \ , \\
&&S_{eff}[q''(\cdot);t+\delta t,t]\approx \frac{m_e(t)}{2\delta t}\delta q''^2-\delta t U_e(\bar q'',t)\ , \\
&& {\mathcal C}_I [q'(\cdot),q''(\cdot);t+\delta t, t]\nonumber \\
&&\hspace{.2in}\approx \tilde {\mathcal K}_I^{(1)}(t)(\bar q'-\bar q'')(\delta q'+\delta q'')  \ ,  \\
&& {\mathcal C}_R[q'(\cdot),q''(\cdot);t+\delta t,t]\approx {\mathcal K}_R^{(0)}(t)(\bar q'-\bar q'')^2 \delta t \nonumber \\
&&\hspace{1in}-\tilde {\mathcal K}_R^{(1)}(t)(\bar q'-\bar q'')(\delta q'-\delta q'')   \nonumber \\
&&\hspace{1in}-\frac{{\mathcal K}_R^{(2)}(t)}{2\delta t} (\delta q'-\delta q'')^2 \ .
\een 
Inserting these expressions into Eqs. (\ref{eq:jeff-def}) and (\ref{eq:sigmae-path}), we find that 
\ben 
&&\tilde \sigma_e(q_f',q_f'';t+\delta t) \approx \frac{m_e(t)}{2\pi\hbar \delta t} \int dq' \int dq'' \tilde \sigma_e(q',q'';t) \nonumber \\
&&\hspace{.2in}\times \exp \Big \{\frac{im_e(t)}{2\hbar\delta t} \Big (\delta q'^2 - \delta q''^2\Big )-\frac{{\mathcal K}_R^{(2)}(t)}{2\hbar\delta t} (\delta q'-\delta q'')^2 \nonumber \\
&&\hspace{.6in}-\frac{i}{\hbar} \tilde {\mathcal K}_I^{(1)}(t) (\bar q'-\bar q'')(\delta q'+\delta q'')\nonumber \\
&&\hspace{.6in}+\frac{1}{\hbar}\tilde {\mathcal K}_R^{(1)}(t) (\bar q'-\bar q'')(\delta q'-\delta q'')\nonumber \\
&&\hspace{.6in} -\frac{\delta t}{\hbar}{\mathcal K}_R^{(0)}(t) (\bar q'-\bar q'')^2  \nonumber \\
&&\hspace{.6in} -\frac{i\delta t}{\hbar} \Big (U_e(\bar q',t) - U_e(\bar q'',t) \Big )\Big \} \ .  \label{eq:rhoe-2}
\een

The remaining steps in deriving a time evolution equation for $\tilde \sigma_e$ from Eq. (\ref{eq:rhoe-2}) are (i) to expand $\tilde \sigma_e(q',q'';t)$ in the integrand around $q_f'$ and $q_f''$, (ii) to perform integrations with respect to 
$\delta q'=q_f'-q'$ and $\delta q''=q_f''-q''$, and (iii) to retain terms up to the order of $\delta t$ only.  The integrations with respect to $\delta q'$ and $\delta q''$ can be done through analytic continuation of the integrands into the complex domain of space followed by normal mode transformation,  which results in standard Gaussian integrations.  Appendix B provides detailed description of all the steps (i)-(iii) of calculations listed above.
The resulting expression, Eq. (\ref{eq:dt-dsigmae-1}) or (\ref{eq:dt-dsigmae-2}), can be summarized as   
\ben
&&\frac{\partial}{\partial t}\tilde  \sigma_e(q',q'';t)=\Big \{ \frac{i\hbar}{2m_e(t)}\Big (\frac{\partial}{\partial q'^2} -\frac{\partial^2}{\partial q''^2}\Big ) \nonumber \\
&&\hspace{.2in} -\frac{i}{\hbar} \Big (U_e(q',t)-U_e(q'',t)\Big) -\frac{\alpha (t)}{\hbar}(q'-q'')^2 \nonumber \\
&&\hspace{.2in} -\frac{\tilde{\mathcal K}_I^{(1)}(t)}{m_e(t)} (q'-q'') \Big (\frac{\partial }{\partial q'} -\frac{\partial}{\partial q''}\Big )  \nonumber \\
&&\hspace{.2in}-i\left (\frac{\tilde {\mathcal K}_R^{(1)}(t)}{m_e(t)} -2 \frac{{\mathcal K}_R^{(2)}(t)}{m_e(t)^2} \tilde {\mathcal K}_I^{(1)}(t)\right) \nonumber \\
&&\hspace{.8in}\times (q'-q'') \Big (\frac{\partial }{\partial q'} +\frac{\partial}{\partial q''}\Big ) \nonumber \\
&&\hspace{.2in}  +\frac{\hbar{\mathcal K}_R^{(2)}(t)}{2m_e(t)^2}\Big (\frac{\partial^2}{\partial q'^2}+\frac{\partial^2}{\partial q''^2}+2\frac{\partial^2}{\partial q' \partial q''}\Big )  \Big\}\tilde \sigma_e(q',q'';t) \ . \nonumber \\ \label{eq:dt-dsigmae}
\een
where $\alpha(t)$ is a real valued function defined by Eq. (\ref{eq:alpha1-def}) or (\ref{eq:alpha2-def}) and can be expressed as follows:
\ben
&&\alpha(t)={\mathcal K}_R^{(0)}(t)-\frac{2}{m_e(t)}\tilde {\mathcal K}_I^{(1)}(t)\Big (\tilde K_R^{(1)}(t) \nonumber \\ 
 &&                                                    \hspace{1.in}-\frac{1}{m_e(t)}{\mathcal K}_R^{(2)}(t)\tilde K_I^{(1)}(t)\Big) \ .
\een

Equivalently, we can express $\tilde \sigma_e(q',q'';t)$ in an operator form as follows: 
\be
\hat {\tilde \sigma}_e(q_g(\cdot);t)= \int dq' \int dq'' |q' \rangle\tilde \sigma_e(q',q'',q_g(\cdot);t)\langle q''|\ , \label{eq:sigmae-op}
\ee
where the dependence on $q_g(\cdot)$ has been shown explicitly.
Then, Eq. (\ref{eq:dt-dsigmae}) can be translated into a time evolution equation for this operator as follows:
\ben
&&\frac{\partial}{\partial t} \hat {\tilde \sigma}_e (q_g(\cdot);t)=-\frac{i}{\hbar} [\hat H_{eff}(t),\hat {\tilde \sigma}_e]-\frac{i}{\hbar}\frac{\tilde {\mathcal K}_I^{(1)}(t)}{m_e(t)}\Big [\hat q,\{\hat p,\hat {\tilde \sigma}_e \}\Big]\nonumber \\
&&\hspace{.2in}+\frac{1}{\hbar} \Big (\frac{\tilde {\mathcal K}_R^{(1)}(t)}{m_e(t)}-2\frac{{\mathcal K}_R^{(2)}(t)}{m_e(t)^2}\tilde {\mathcal K}_I^{(1)}(t)\Big) \Big [\hat q,[\hat p,\hat {\tilde \sigma}_e]\Big]\nonumber \\
&&\hspace{.2in}-\frac{\alpha(t)}{\hbar} \Big [\hat q,[\hat q,\hat {\tilde \sigma}_e]\Big] -\frac{{\mathcal K}_R^{(2)}(t)}{2\hbar m_e(t)^2}\Big [\hat p,[\hat p, \hat {\tilde \sigma}_e]\Big]\ ,  \label{eq:dt-dsigmae-op}
\een
where 
\be
\hat H_{eff}(t)=\frac{\hat p^2}{2m_e(t)}+U_{e}(\hat q,q_g(\cdot),t)\ ,
\ee  
with $U_e(\hat q,q_g(\cdot),t)$ defined by Eq. (\ref{eq:ue_def}). Note that the effects of the ground state bath appear only in the effective time dependent potential $U_e(\hat q,q_g(\cdot),t)$.  A phase space representation for Eq. (\ref{eq:dt-dsigmae-op}), which is a more conventional form of QFPE, can be found by applying the Wigner transformation\cite{hillery-pr106} to Eq. (\ref{eq:dt-dsigmae-op}).

Equation (\ref{eq:dt-dsigmae-op}) is in the Dekker form\cite{diosi-pa199,dekker-pr80} unlike the original CL's QFPE.\cite{caldeira-pa121}  Thus, the Lindblad's condition of positive definiteness\cite{lindblad-cmp48} can be satisfied for appropriate range of physical variables.  To this end, more detailed analysis of each term is necessary for a specific form of the spectral density chosen. 

\section{Results for Ohmic Spectral Density with Drude Cutoff}
Let us consider the case where the excited state bath spectral density, Eq. (\ref{eq:eta_e}), is given by the Ohmic spectral density with Drude cutoff as follows:
\be
\eta_e(\omega)=2m\gamma_e \frac{\omega_c^2\omega}{\omega^2+\omega_c^2}=2 m\gamma_s \omega_c^2 \frac{\omega/\omega_c}{(\omega/\omega_c)^2+1} \ ,
\ee
where $\gamma_e$ is the friction constant in the excited electronic state and $\gamma_s=\gamma_e/\omega_c$ is a dimensionless scaled version of the same friction constant. 
For the above spectral density, $\kappa_e$, defined by Eq. (\ref{eq:lam_e}), has the following form: 
\be
\kappa_e=2m\gamma_e\omega_c=2m\gamma_s\omega_c^2\ ,
\ee
and the imaginary component of the bath correlation function, Eq. (\ref{eq:eta-it}), can be expressed as 
\be
\tilde \eta_{e,I}(t)=m\gamma_e\omega_c^2 e^{-\omega_c t}\  .
\ee
Then, it is straightforward to show that ${\mathcal K}_I^{(n)}(t)$ defined by Eq. (\ref{eq:kint}), for $n=0-2$, can be expressed as
\ben 
&&{\mathcal K}_I^{(0)}(t)=m\gamma_e\omega_c F_0(t_s) \ ,\label{eq:ki0t-ohmic}\\
&&{\mathcal K}_I^{(1)}(t)=m\gamma_e F_1(t_s) \ , \label{eq:ki1t-ohmic}\\
&&{\mathcal K}_I^{(2)}(t)=2m\gamma_{s} F_2(t_s) \ , \label{eq:ki2t-ohmic}
\een 
where $t_s=\omega_c t$ is a scaled time, and 
\ben 
&&F_0(t_s)=1-e^{-t_s}\ , \\
&&F_1(t_s)=1-(1+t_s)e^{-t_s}\ , \\
&&F_2(t_s)=1-\Big (1+t_s+\frac{t_s^2}{2}\Big)e^{-t_s} \ , 
\een
which all approach $1$ in the limit of $t_s\rightarrow \infty$.  All of these three functions are monotonically increasing and positive for $t_s >0$.
Combining Eq. (\ref{eq:ki1t-ohmic}) and the time derivative of Eq. (\ref{eq:ki2t-ohmic}), we also obtain the expression for $\tilde {\mathcal K}_1^{(1)}(t)$, defined by Eq. (\ref{eq:kitil1}), as follows:
\be
\tilde {\mathcal K}_I^{(1)} (t)=m\gamma_e \tilde F_1(t_s) \ ,
\ee 
where
\be
 \tilde F_1(t_s)=1-\Big (1+t_s-\frac{t_s^2}{2}\Big)e^{-t_s} \ .
 \ee
Note that $\tilde F_1(t_s)$ also approaches $1$ in the limit  of $t_s\rightarrow \infty$ and is positive for $t_s >0$. 

Taking the ratio of the effective time dependent mass $m_e(t)$, Eq. (\ref{eq:me_def}), to the actual mass $m$, let us introduce
\be
r_m(t)=\frac{m_e(t)}{m}= 1-2\gamma_s F_2(t_s) \ ,
\ee 
which approaches $1-2\gamma_s$ in the limit of $t\rightarrow \infty$.
  
The real component of the bath correlation function, Eq. (\ref{eq:eta-rt}), can be evaluated employing the well-known Matsubara expansion of  $\coth (x)$ and $\cot(x)$, and is expressed as
\ben
&&\tilde \eta_{e,R}(t)=m\gamma_e\omega_c^2\cot (\frac{\beta\hbar \omega_c}{2})e^{-\omega_c t}\nonumber \\
&&\hspace{.4in}+\frac{4 m\gamma_e\omega_c^2}{\beta\hbar}\sum_{n=1}^\infty \frac{\omega_n}{(\omega_n^2-\omega_c^2)} e^{-\omega_n t}  \ ,
\een 
where $\omega_n=2\pi n/(\beta\hbar)$.
Then, it is straightforward to show that ${\mathcal K}_R^{(n)}(t)$ defined by Eq. (\ref{eq:krnt}), for $n=0-2$, can be expressed as 
\ben
&&{\mathcal K}_R^{(0)}(t)=m\gamma_e \omega_c G_0(\beta_s,t_s)\ ,\\
&&{\mathcal K}_R^{(1)}(t)=m\gamma_e G_1(\beta_s,t_s) \ , \label{eq:kr1t-ohmic} \\
&&{\mathcal K}_R^{(2)}(t)=m\gamma_s G_2(\beta_s,t_s) \ , \label{eq:kr2t-ohmic}
\een 
where $\beta_s=\beta\hbar \omega_c$, a dimensionless and scaled inverse temperature, and 
\ben
&&G_0(\beta_s,t_s)= \cot\left (\frac{\beta_s}{2} \right )F_0(t_s)\nonumber \\
&&\hspace{.6in}+\frac{4}{\beta_s}\sum_{n=1}^\infty \frac{F_0(2\pi nt_s/\beta_s)}{(2\pi n/\beta_s)^2-1} \ ,\\
&&G_1(\beta_s,t_s)=\cot\left (\frac{\beta_s}{2} \right )F_1(t_s)\nonumber \\
&&\hspace{.6in}+\frac{4}{\beta_s}\sum_{n=1}^\infty \frac{F_1(2\pi n t_s/\beta_s)}{(2\pi n/\beta_s)((2\pi n/\beta_s)^2-1)} \ , \\
&&G_2(\beta_s,t_s)=\cot\left (\frac{\beta_s}{2} \right )F_2(t_s)\nonumber \\
&&\hspace{.6in}+\frac{4}{\beta_s}\sum_{n=1}^\infty \frac{F_2(2\pi n t_s/\beta_s)}{(2\pi n/\beta_s)^2((2\pi n/\beta_s)^2-1)} \ .  
\een 
In addition, combination of Eq. (\ref{eq:kr1t-ohmic}) and the time derivative of Eq. (\ref{eq:kr2t-ohmic}) leads to the expression for $\tilde {\mathcal K}_R^{(1)}(t)$, defined by Eq. (\ref{eq:krtil1}), as follows:
 \ben
&&\tilde {\mathcal K}_R^{(1)}(t)=m\gamma_e \tilde G_1(\beta_s;t_s) \ ,
\een
where
 \ben
&&\tilde G_1(\beta_s;t_s)=\cot \left ( \frac{\beta_s}{2} \right )\tilde F_1 (t_s)\nonumber \\
&&\hspace{.6in}+\frac{4}{\beta_s}\sum_{n=1}^\infty \frac{\tilde F_1 (2\pi n t_s/\beta_s)}{(2\pi n/\beta_s)((2\pi n/\beta_s)^2-1)} \ .
\een

Employing the above expressions, Eq. (\ref{eq:dt-dsigmae-op}) for the present spectral density can be expressed as 
\ben
&&\frac{\partial}{\partial t} \hat {\tilde \sigma}_e (q_g(\cdot);t)=-\frac{i}{\hbar} [\hat H_{eff}(t),\hat {\tilde \sigma}_e]\nonumber \\
&&-\frac{i\gamma_e}{\hbar} \Gamma (t)\Big [\hat q,\{\hat p,\hat {\tilde \sigma}_e \}\Big]+\frac{\gamma_e}{\hbar} {\mathcal R}_{pq} (t) \Big [\hat q,[\hat p,\hat {\tilde \sigma}_e]\Big]\nonumber \\
&&-\frac{m\gamma_e^2}{\hbar} {\mathcal R}_{pp}(t) \Big [\hat q,[\hat q,\hat {\tilde \sigma}_e]\Big] -\frac{{\mathcal R}_{qq}(t)}{m \hbar }\Big [\hat p,[\hat p, \hat {\tilde \sigma}_e]\Big]\ ,  \label{eq:dt-dsigmae-op-ohmic} 
\een
where 
\ben
&&\Gamma (t)=\frac{\tilde F_1(t_s)}{r_m(t)} \ ,\\
&&{\mathcal R}_{pq}(t)=\frac{1}{r_m(t)}\tilde G_1(\beta_s,t_s) -2\frac{\gamma_s}{r_m^2} G_2(\beta_s,t_s)\tilde F_1(t_s)\ , \nonumber  \\ \\
&&{\mathcal R}_{qq}(t)=\frac{\gamma_s}{2 r_m(t)^2} G_2(\beta_s,t_s) \ , \\
&&{\mathcal R}_{pp}(t)=\frac{1}{\gamma_s}G_0(\beta_s,t_s)- 2 \Gamma(t)\tilde G_1(\beta_s,t_s) \nonumber \\
&&\hspace{.5in} +2\gamma_s \Gamma(t)^2 G_2(\beta_s,t_s)  \ .
\een

As mentioned in the previous section, Eq. (\ref{eq:dt-dsigmae-op-ohmic}) is in the Dekker form\cite{diosi-pa199,dekker-pr80} and satisfies the Lindblad's positive definiteness condition\cite{lindblad-cmp48} given that the following inequality holds.
\be
D(t)\equiv 4{\mathcal R}_{pp} (t){\mathcal R}_{qq}(t)-{\mathcal R}_{pq}(t)^2-\Gamma(t)^2 >0 \ . \label{eq:det}
\ee
\begin{figure}
\includegraphics[width=3.in]{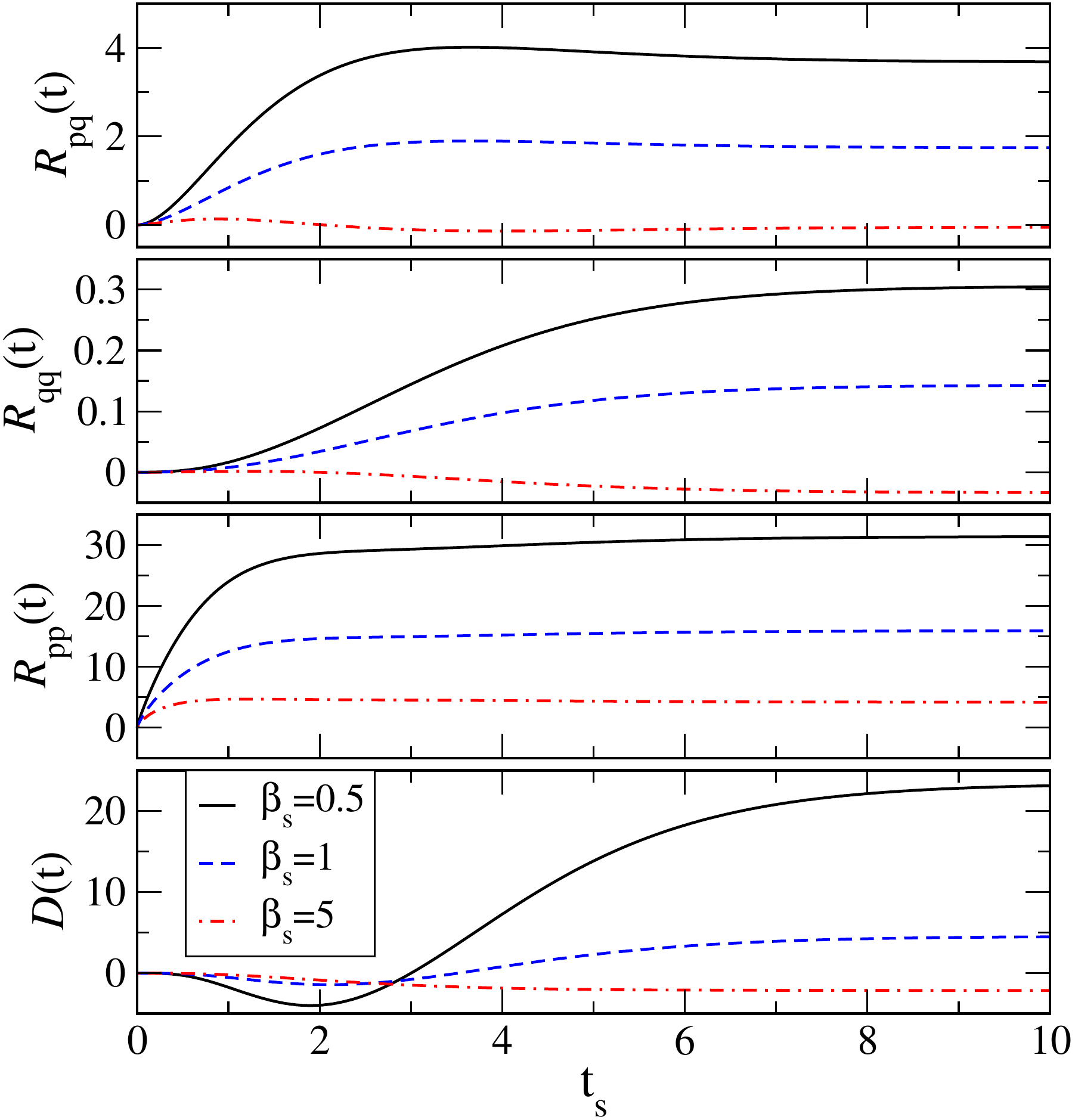}
\caption{Values of ${\mathcal R}_{pq}(t)$, ${\mathcal R}_{qq}(t)$, ${\mathcal R}_{pp}(t)$, and $D(t)$ versus $t_s=\omega_c t$ for $\beta_s=0.5$, $1$, and $5$.  The value of $\gamma_s=0.1$.} 
\end{figure}

Figure 1 shows calculated results of ${\mathcal R}_{pq}(t)$, ${\mathcal R}_{qq}(t)$, ${\mathcal R}_{pp}(t)$, and $D(t)$ for three cases of $\beta_s=0.5$, $1$, and $5$.  A small value of $\gamma_s=0.1$ was chosen, for which the weak damping condition of $\gamma_s G_2(\beta_s,t_s) <1-2\gamma_s F_2(t_s)$ is satisfied throughout the entire time.   Except for the case of very low temperature, $\beta_s=5$, all the values of ${\mathcal R}_{qq}(t)$, ${\mathcal R}_{qq}(t)$, and ${\mathcal R}_{pp}(t)$ remain positive.  Although $D(t)$ becomes negative transiently in early stage, its steady state limits are positive for $\beta_s=0.1$ and $1$.   This shows that the non positivity condition of CL's QFPE\cite{caldeira-pa121} can be fixed by avoiding inconsistent use of Markovian approximation, confirming the analysis by Diosi.\cite{diosi-pa199} On the other hand, for $\beta_s=5$, $D(t)$ becomes negative for all values of $t>0$.  The main contribution to this negative value comes from that of ${\mathcal R}_{pp}(t)$, and indicates that the second order approximation for the real part of the bath correlation function is not valid at this temperature due to nonlocality of the quantum dynamics in time.     

\section{Conclusion}

The present work has provided a derivation of a GQFPE by extending CL's path integral approach.\cite{caldeira-pa121}  
The only assumption used in this derivation is that the bath correlation functions are short ranged in time so that the second order expansions of trajectories  within the integrands in the exponents of the influence functional are well justified.  This seems to be the most general assumption one can make in order to convert the double time integrations in the exponents of the influence functional into single time integrations, from which a time evolution equation can be derived.  
Thus, the resulting GQFPE, Eq. (\ref{eq:dt-dsigmae-op}), may serve as a general form that can include various known QFPEs as special cases.  In addition, Eq. (\ref{eq:dt-dsigmae-op}) can also serve as a new and useful means to describe photo-induced quantum relaxation processes  beyond typical high temperature and weak coupling limits, and thus will serve as a more satisfactory theoretical tool to study wider range of photoinduced electron and proton transfer processes.

The importance of the general form of the GQFPE, Eq. (\ref{eq:dt-dsigmae-op}), is that it has not been constructed phenomenologically, but rather derived from a well defined Hamiltonian. Thus, it offers detailed microscopic expressions for all the terms entering the equation in terms of the parameters defining the Hamiltonian. This makes it possible to examine the validity of the assumptions underlying the derivation {\it a posteriori}, for a given Hamiltonian and bath spectral density.  Most of all, because Eq. (\ref{eq:dt-dsigmae-op}) is in the Dekker form,\cite{diosi-pa199,dekker-pr80} the condition of positive definiteness can be tested explicitly.

For one of the most well-known spectral densities, the Ohmic spectral density with Drude cutoff, all terms in the GQFPE have been calculated explicitly in Sec. III.  The results demonstrated in Fig. 1 show that the GQFPE indeed is well defined and its steady state limit satisfies the Lindblad's condition\cite{lindblad-cmp48} for reasonable physical situation.  This confirms that the violation of positive definiteness results from an inconsistent application of the Markovian approximation or the breakdown of time locality in the quantum dynamics as is typical at very low temperature regime.  

There remain some subtle issues that need to be clarified in the future.  For example, the physical implication of the boundary terms in Eq. (\ref{eq:boundary-term}) should be understood better.   In addition,  the expressions of terms in Eq. (\ref{eq:dt-dsigmae-op}) and Eq. (\ref{eq:dt-dsigmae-op-ohmic}) show that the detailed manner of the high frequency cutoff, even for the Ohmic spectral density, makes significant contribution to the final form of the equation. This is consistent with a previous analysis based on a fourth order quantum master equation.\cite{jang-jcp116} This also shows the possibility that Eq. (\ref{eq:dt-dsigmae-op}) can serve as a useful theoretical tool to examine the type of spectral density and physical conditions for which a time local QFPE can be established.

The methodology of present  work can easily be extended to multidimensional system and nonadiabatic cases.  Consideration of these cases will be another subject of future theoretical investigation.  Finally, further test of the GQFPE against exactly solvable models\cite{hu-prd45,haake-pra32,karrlein-pre55,ford-prl82} and virtually exact calculation approaches\cite{yan-arpc56,tanimura-jcp142} remain as important future tasks.  Outcomes of these studies will help provide ultimate validation of the GQFPE of this work and understanding of the extent to which a closed form time local equation can be used to describe nonequilibrium and non-Markovian quantum dynamical processes.

\acknowledgements

The author acknowledges the support for this research from  the National Science Foundation  (CHE-1362926), the Office of Basic Energy Sciences, Department of Energy (DE-SC0001393),  and the Camille Dreyfus Teacher Scholar Award. 

\appendix

\section{Nonequilibrium Influence Functional}
The path integral expression for Eq. (\ref{eq:sigmae-def}) and an appropriate expression 
for the influence functional, which are well known,\cite{weiss} are derived here for the sake of completeness.  
For the real time propagator, $e^{-iH_et/\hbar}$, the path integral expression is given by
\ben
&&e^{-i\hat H_e t/\hbar}= \int dq_i \int dq_f \int dx_i\int dx_f \ |q_f,x_f\rangle \langle q_i,x_i| \nonumber \\
&&\times \int_{q_i}^{q_f} {\mathcal D}q(\cdot)\int_{x_i}^{x_f} {\mathcal D} x(\cdot) e^{\frac{i}{\hbar}S_e[q(\cdot);t]+\frac{i}{\hbar}S_{eb}[x(\cdot),q(\cdot);t]}\comma \nonumber \\ \label{eq:rpath1}
\een 
where $x\equiv (x_1,\cdots,x_\alpha,\cdots)$ and $|q,x\rangle \equiv |q\rangle \prod_\alpha |x_\alpha\rangle$.  $S_e$ and $S_{eb}$ are real time actions of the excited state, respectively given by
\be
S_e[q(\cdot);t]=\int_0^t dt_1 \left (\frac{m}{2}\dot q(t_1)^2-V_e(q(t_1))-\frac{\kappa_e}{2} q(t_1)^2\right) \comma
\ee
with 
\be
\kappa_e=\sum_\alpha \frac{c_{\alpha,e}^2}{m_\alpha \omega_\alpha^2}=\frac{2}{\pi}\int_0^\infty d\omega \frac{\eta_e(\omega)}{\omega} \ , \label{eq:lam_e}
\ee 
and 
\ben
&&S_{eb}[x(\cdot), q(\cdot);t]=\sum_\alpha\int_0^t dt_1 \left (\frac{m_\alpha}{2}\dot x_\alpha(t_1)^2\right . \nonumber \\
&&\hspace{.5in}\left . -\frac{m_\alpha \omega_\alpha^2}{2}x_\alpha(t_1)^2 +c_{\alpha,e}q(t_1)x_\alpha(t_1)\right ) \perd
\een
On the other hand, for $e^{-\beta \hat H_g}$, the path integral expression is given by
\ben
&&e^{-\beta \hat H_g} = \int dq' \int dq'' \int d x' \int d x'' |q',x'\rangle \langle q'',x''| \nonumber \\
&&\times \int_{q'}^{q''} {\mathcal D}q(\cdot) \int_{x'}^{x''}{\mathcal D}x(\cdot) e^{- \frac{1}{\hbar}S_g^E[q(\cdot);\beta\hbar]-\frac{1}{\hbar}S_{gb}^E[q(\cdot),x(\cdot);\beta\hbar]}\comma  \nonumber \\ \label{eq:ipath1}
\een
where $S_g^E$ and $S_{gb}^E$ are Euclidean actions, respectively given by
\ben
&&S_g^E[q(\cdot);\beta\hbar]=\nonumber \\
&&\hspace{.2in}\int_0^{\beta\hbar}d\tau \left (\frac{m}{2}\dot q(\tau)^2+V_g(q(\tau))+\frac{\kappa_g}{2}f_g(r(\tau))^2\right)\comma \nonumber \\
\een
with $\kappa_g=\sum_\alpha c_{\alpha,g}^2/(m_\alpha \omega_\alpha^2)$, and
\ben
&&S_{gb}^E[x(\cdot), q(\cdot);\beta\hbar]=\sum_\alpha\int_0^t dt_1 \left (\frac{m_\alpha}{2}\dot x_\alpha(t_1)^2\right .\nonumber \\
&&\hspace{.2in}\left .+\frac{m_\alpha \omega_\alpha^2}{2}x_\alpha(t_1)^2 -c_{\alpha,g}q(t_1)x_\alpha(t_1)\right ) \perd
\een

The path integral expression for $\hat \sigma_e(t)$ can be obtained by inserting 
Eq. (\ref{eq:rpath1}), its complex conjugate, and Eq. (\ref{eq:ipath1}),  
into Eq. (\ref{eq:sigmae-def}).  By performing explicit path integration over the bath degrees of freedom, one 
can show that Eq. (\ref{eq:rpath1}) reduces to 
\ben
&&e^{-i\hat H_e t/\hbar}=\int dq_i \int dq_f \int dx_i \int dx_f |q_f,x_f\rangle \langle q_i,x_i| \nonumber \\ 
&&\hspace{.2in}\times \int_{q_i}^{q_f} {\mathcal D}q(\cdot)  e^{ \frac{i}{\hbar}S_e[q(\cdot);t]}\prod_{\alpha}T_{e,\alpha}[q(\cdot);x_{\alpha,f},x_{\alpha,i},t] \comma \nonumber \\ \label{eq:rpath2}
\een
where
\ben
&&T_{e,\alpha}[r(\cdot);x_{\alpha,f},x_{\alpha,i};t]=\left (\frac{m_\alpha \omega_\alpha}{2\pi i\hbar \sin (\omega_\alpha t)}\right)^{1/2}\nonumber \\
&&\times \exp \Bigg \{\frac{i}{\hbar}\Big [ \frac{m_\alpha \omega_\alpha}{2\sin (\omega_\alpha t)} \left ( x_{\alpha,f}^2\cos(\omega_\alpha t)\right . \nonumber \\
&&\hspace{.6in}\left . +x_{\alpha,i}^2 \cos(\omega_\alpha t)-2x_{\alpha,f} x_{\alpha,i} \right)  \nonumber \\
&&\hspace{.4in}+\frac{c_{\alpha,e}x_{\alpha,f}}{\sin (\omega_\alpha t)}\int_0^t dt_1 \sin (\omega_\alpha t_1)q(t_1)\nonumber \\
&&\hspace{.4in}+\frac{c_{\alpha,e}x_{\alpha,i}}{\sin (\omega_\alpha t)}\int_0^t dt_1 \sin (\omega_\alpha(t- t'))q(t_1) \nonumber \\
&& \hspace{.4in}-\frac{c_{\alpha,e}^2}{m_\alpha \omega_\alpha \sin (\omega_\alpha t)} \int_0^t dt_1\int_0^{t_1} dt_2 \sin(\omega_\alpha (t-t_1))\nonumber \\
&&\hspace{.8in}\times \sin (\omega_\alpha t_2) q(t_1) q(t_2) \Big]\Bigg\} \perd
\een
Similarly, Eq. (\ref{eq:ipath1}) can be shown to be
\ben
&&e^{-\beta \hat H_g} = \int dq' \int dq'' \int d x' \int d x'' |q',x'\rangle \langle q'',x''| \nonumber \\
&&\hspace{.2in}\times \int_{r'}^{r''} {\mathcal D}q(\cdot) e^{- \frac{1}{\hbar}S_g^E[q(\cdot);\beta\hbar]}\prod_{\alpha}T_{g,\alpha}^E[q(\cdot);x'_\alpha,x''_\alpha,\beta\hbar] \comma \nonumber \\ \label{eq:ipath2}
\een
where 

\ben
&&T_{g,\alpha}^E[q(\cdot);x'_\alpha,x''_\alpha,\beta\hbar]=\left (\frac{m_\alpha \omega_\alpha}{2\pi\hbar \sinh (\omega_\alpha \beta\hbar)}\right)^{1/2} \nonumber \\
&&\times \exp \Bigg \{-\frac{1}{\hbar}\Big [ \frac{m_\alpha \omega_\alpha}{2\sinh (\omega_\alpha \beta\hbar)} \left ( x_\alpha'^2\cosh(\omega_\alpha \beta\hbar) \right . \nonumber  \\
&&\hspace{.6in}\left . +x_\alpha''^2 \cosh (\omega_\alpha \beta\hbar)-2x'_\alpha x''_\alpha \right)  \nonumber \\
&&\hspace{.4in}-\frac{c_{\alpha,g}x'_\alpha}{\sinh (\omega_\alpha \beta\hbar)}\int_0^{\beta\hbar} d\tau \sinh (\omega_\alpha \tau)q_g(\tau)\nonumber \\
&&\hspace{.4in}-\frac{c_{\alpha,g}x_\alpha''}{\sinh (\omega_\alpha \beta\hbar)}\int_0^{\beta\hbar} d\tau \sinh (\omega_\alpha(\beta\hbar-\tau))q_g(\tau) \nonumber \\
&&\hspace{.4in} -\frac{c_{\alpha,g}^2}{m_\alpha \omega_\alpha \sinh (\omega_\alpha \beta\hbar)} \int_0^{\beta\hbar} d\tau \int_0^{\tau} d\tau_1 \nonumber \\
&&\hspace{.6in} \times \sinh(\omega_\alpha (\beta\hbar-\tau)) \sinh (\omega_\alpha \tau_1)q_g(\tau)q_g(\tau_1) \Big ] \Bigg \} \perd\nonumber \\
\een

With the use of Eq. (\ref{eq:rpath2}), its complex conjugate, and 
Eq. (\ref{eq:ipath1}) in Eq. (\ref{eq:sigmae-def}), the reduced density operator can be expressed as 

\ben
&&\hat \sigma_e(t)=\frac{1}{Z_g}\int dq_f'\int dq_f''\ |q_f'\rangle\langle q_f''| \nonumber \\
&&\times \int dq_i'\int dq_i''\int dx_f \int dx_i'\int dx_i'' \nonumber \\ 
&&\times \int_{q_i'}^{q_f'} {\mathcal D} q'(\cdot)\int_{q_i''}^{q_f''} {\mathcal D}q''(\cdot) \int_{q_i'}^{q_i''} \dpt q_g(\cdot) \exp\Bigg\{\frac{i}{\hbar}S_e[q'(\cdot);t] \nonumber \\
&&\hspace{.8in}-\frac{i}{\hbar}S_e[q''(\cdot);t]-\frac{1}{\hbar}S_g^E[q_g(\cdot);\beta\hbar]\Bigg\} \nonumber \\
&&\hspace{.2in}\times \prod_\alpha \Bigg \{T_{e,\alpha}[q'(\cdot);x_{\alpha,f},x_{\alpha,i}';t]T_{e,\alpha}^*[q''(\cdot);x_{\alpha,f},x_{\alpha,i}'';t]\nonumber \\
&&\hspace{.4in}\times T_{g,\alpha}^E[q_g(\cdot);x_\alpha',x_\alpha'',\beta\hbar]\Bigg\}\perd \label{eq:pt2}
\een
Performing integrations over $x_f$, $x_i'$, and $x_i''$ leads to Eqs. (\ref{eq:inf-def}) and (\ref{eq:sigmae-1}) with the following  definitions of its exponents.
\ben
W_{e,I}(t)&=&\int_0^t d t_1\int_0^{t_1} dt_2 \int_0^\infty \frac{d\omega}{\pi} \eta_e (\omega)\sin (\omega (t_1-t_2)) \nonumber \\
&&\times (q'(t_1)-q''(t_1))(q'(t_2)+q''(t_2)) \ , \nonumber \\ \label{eq:w1t}\\
W_{e,R}(t)&=&\int_0^t dt_1\int_0^{t_1} dt_2 \int_0^\infty \frac{d\omega}{\pi} \eta_e(\omega)\nonumber \\
&&\times \coth(\frac{\omega\beta\hbar}{2})\cos (\omega (t_1-t_2))\nonumber \\
&&\times (q'(t_1)-q''(t_1))(q'(t_2)-q''(t_2)) \ , \nonumber \\ \label{eq:w2t}\\
W_c(t)&=&\int_0^tdt_1\int_0^{\beta\hbar} d\tau \int_0^\infty \frac{d \omega}{\pi} \eta_c(\omega)\nonumber \\ 
&&\times  \frac{\cosh(\omega(\tau-it_1-\beta\hbar/2))}{\sinh (\omega\beta\hbar/2)}\nonumber \\
&&\times  (q'(t_1)-q''(t_1))q_g(\tau)\ , \nonumber \\ \label{eq:w3t} \\
W_g(\beta\hbar)&=&\int_0^{\beta \hbar}d\tau\int_0^{\tau}d\tau_1 \int_0^{\infty}\frac{d\omega}{\pi} \eta_g(\omega)\nonumber \\&&\times \frac{\cosh(\omega(\tau-\tau_1-\beta\hbar/2))}{\sinh (\omega\beta\hbar/2)}q_g(\tau_1)q_g(\tau)\ . \nonumber \\ \label{eq:w4t}
\een

\section{Derivation of Eq. (\ref{eq:dt-dsigmae})}

In the integrand of Eq. (\ref{eq:rhoe-2}), consider the following term:
\ben
&&Q(\delta q',\delta q'')=\frac{m_e}{2\hbar \delta t}\Big \{i\delta q'^2-i\delta q''^2 \nonumber \\
&&\hspace{.8in}-\frac{1}{m_e} {\mathcal K}_R^{(2)}(t) \left (\delta q'-\delta q''\right)^2 \Big\}\ .\label{eq:q_exp}
\een
This term can be diagonalized into a sum of two quadratic terms by using complex-valued coordinates, the choice of which depends on the magnitude of $ {\mathcal K}_R^{(2)}(t)/m_e$ as described below. In the above expressions, the time dependence of $m_e(t)$ has not been shown explicitly and will remain so throughout this section.

\subsubsection{Case for $0< {\mathcal K}_R^{(2)}(t)/m_e < 1$}

For this case, we can introduce $\mu_1 (t)$ such that 
\be
\sin (\mu_1(t))=\frac{{\mathcal K}_R^{(2)}(t)}{m_e}  \ ,\label{eq:gamma1_def}
\ee
where $0 < \mu_1(t) < \pi/2$.
Solving the eigenvalue problem for the quadratic form, Eq. (\ref{eq:q_exp}), it is straightforward to find out the following two normal modes defined in the complex domain.
\ben
&&u'=\frac{1}{\sqrt{\cos\mu_1}}\left ( \cos(\frac{\mu_1}{2}) \delta q'-i\sin (\frac{\mu_1}{2})\delta q''\right)\ , \\
&&u''=\frac{1}{\sqrt{\cos\mu_1}} \left (i\sin (\frac{\mu_1}{2}) \delta q'+\cos(\frac{\mu_1}{2})\delta q''\right)\ .
\een
Equivalently, $\delta q'$ and $\delta q''$ can be expressed in terms of $u'$ and $u''$ as follows:
\ben
&&\delta q'=\frac{1}{\sqrt{\cos\mu_1}}\left ( \cos(\frac{\mu_1}{2}) u'+i\sin (\frac{\mu_1}{2})u''\right)\ , \label{eq:delq'} \\
&&\delta q''=\frac{1}{\sqrt{\cos\mu_1}} \left (-i\sin (\frac{\mu_1}{2}) u'+\cos(\frac{\mu_1}{2})u''\right)\ . \label{eq:delq''}
\een
Inserting these expressions into Eq. (\ref{eq:q_exp}), we find that
\ben
Q(\delta q',\delta q'')&=&\frac{m_e}{2\hbar \delta t}\big \{(i\cos\mu_1 -\sin \mu_1)u'^2\nonumber \\
&&\hspace{.2in}+(-i\cos\mu_1-\sin\mu_1)u''^2 \big \} \nonumber \\
&=&\frac{m_e}{2\hbar \delta t}\left \{ ie^{i\mu_1}u'^2-ie^{-i\mu_1}u''^2\right\}\ .
\een
Then, Eq. (\ref{eq:rhoe-2}) can be expressed as 
\ben 
&&\tilde \sigma_e(q_f',q_f'';t+\delta t) \approx \frac{m_e}{2\pi\hbar \delta t } \int du' \int du'' \tilde \sigma_e(q',q'';t)\nonumber \\
&&\times  \exp \Big \{\frac{im_e}{2\hbar\delta t} \Big (e^{i\mu_1}u'^2 - e^{-i\mu_1} u''^2\Big )\nonumber \\
&&\hspace{.4in} -\frac{i}{\hbar} \tilde {\mathcal K}_I^{(1)}(t) (\bar q'-\bar q'')\frac{1}{\sqrt{\cos \mu_1}}(e^{-i\mu_1/2}u'+e^{i\mu_1/2}u'') \nonumber \\
&&\hspace{.4in}+\frac{1}{\hbar}\tilde {\mathcal K}_R^{(1)}(t) (\bar q'-\bar q'')\frac{1}{\sqrt{\cos{\mu_1}}}(e^{i\mu_1/2}u'-e^{-i\mu_1/2}u'') \nonumber \\
&&\hspace{.4in}-\frac{\delta t}{\hbar}{\mathcal K}_R^{(0)}(t) (\bar q'-\bar q'')^2  \nonumber \\
&&\hspace{.4in} -\frac{i\delta t}{\hbar} \Big (U_e(\bar q',t) - U_e(\bar q'',t) \Big )\Big \} \ .\label{eq:rhoe-3}
\een
Let us introduce 
\ben
&&\xi'=u'-\frac{i\delta t e^{-i\mu_1}}{m_e\sqrt{\cos\mu_1}} (\bar q'-\bar q'')\nonumber \\
&&\hspace{.4in}\times \left (\tilde {\mathcal K}_R^{(1)}(t)e^{i\mu_1/2}-i\tilde {\mathcal K}_I^{(1)}(t)e^{-i\mu_1/2} \right ) \ , \label{eq:xi'}\\
&&\xi''=u''-\frac{i\delta t e^{i\mu_1}}{m_e\sqrt{\cos\mu_1}} (\bar q'-\bar q'')\nonumber \\
&&\hspace{.4in}\times \left (\tilde {\mathcal K}_R^{(1)}(t)e^{-i\mu_1/2}+i\tilde {\mathcal K}_I^{(1)}(t)e^{i\mu_1/2} \right )\ . \label{eq:xi''}
\een
Then, the squares of  the exponents in Eq. (\ref{eq:rhoe-3}) can be completed as follows:
 \ben 
&&\tilde \sigma_e(q_f',q_f'';t+\delta t) \approx \frac{m_e}{2\pi\hbar \delta t }\int d\xi' \int d\xi'' \tilde \sigma_e(q',q'';t) \nonumber \\
&&\times  \exp \Big \{\frac{im_e}{2\hbar\delta t} \Big (e^{i\mu_1}\xi'^2 - e^{-i\mu_1} \xi''^2\Big )\nonumber \\
&&\hspace{.4in}-\frac{\delta t}{\hbar}\alpha (t) (\bar q'-\bar q'')^2 \nonumber \\
&&\hspace{.4in} -\frac{i\delta t}{\hbar} \Big (U_e(\bar q',t) - U_e(\bar q'',t) \Big ) \Big \} \ ,\label{eq:rhoe-4}
\een
where
\be
\alpha (t) ={\mathcal K}_R^{(0)}(t)-\frac{2}{m_e}\tilde {\mathcal K}_I^{(1)}(t)\Big (\tilde K_R^{(1)}(t)-\sin \mu_1 \tilde K_I^{(1)}(t)\Big)\ . \label{eq:alpha1-def}
\ee

In Eq. (\ref{eq:rhoe-4}), $q'=q_f'-\delta q'$ and $q''=q_f''-\delta q"$ can be expressed as
\ben
&&q'=q_f'-\frac{1}{\sqrt{\cos\mu_1}} \left ( \cos \left (\frac{\mu_1}{2}\right)\xi'+i\sin \left (\frac{\mu_1}{2}\right)\xi''\right ) \nonumber \\
&&\hspace{.4in}-\frac{\delta t}{m_e} {\mathcal K}_c(t) (\bar q'-\bar q'')\ , \\
&&q''=q_f''-\frac{1}{\sqrt{\cos\mu_1}} \left ( \cos \left (\frac{\mu_1}{2}\right)\xi''-i\sin \left (\frac{\mu_1}{2}\right)\xi'\right )\nonumber \\
&&\hspace{.4in} +\frac{\delta t}{m_e}{\mathcal K}_c^*(t)(\bar q'-\bar q'') \ ,
\een
where Eqs. (\ref{eq:delq'}), (\ref{eq:delq''}), (\ref{eq:xi'}), and (\ref{eq:xi''}) have been used and 
\be
{\mathcal K}_c(t)=\tilde {\mathcal K}_I^{(1)}(t)+i\left (\tilde {\mathcal K}_R^{(1)}(t)-2\sin \mu_1 \tilde {\mathcal K}_I^{(1)}(t)\right) \ .
\ee

Inserting the above expressions into the arguments of $\tilde \sigma_e(q',q'';t)$ in Eq. (\ref{eq:rhoe-3}) and expanding the integrand up to the second order of $\xi'$ and $\xi''$,  we find the following expression: 
\ben
&&\tilde \sigma_e(q_f',q_f'';t+\delta t) \approx \nonumber \\
&&\hspace{.1in} \tilde \sigma_e(q_f'-\frac{\delta t  {\mathcal K}_c(t)}{m_e\cos\mu_1} (\bar q'-\bar q''),q_f''+\frac{\delta t {\mathcal K}_c^*(t)}{m_e\cos\mu_1} (\bar q'-\bar q'');t) \nonumber \\
&&\hspace{.1in}\times \exp\left \{-\frac{i\delta t}{\hbar} (U_e(\bar q',t) - U_e(\bar q'',t)) -\frac{\delta t}{\hbar} \alpha(t)(\bar q'-\bar q'')^2\right\} \nonumber \\
&&\hspace{.1in}+\frac{m_e}{4\pi\hbar \delta t\cos \mu_1} \int d\xi'\int d\xi''  \nonumber \\
&&\hspace{.5in}\times \exp\left \{\frac{im_e}{2\hbar\delta t}\left( e^{i\mu_1}\xi'^2-e^{-i\mu_1}\xi''^2\right)\right\} \nonumber \\ 
&&\hspace{.5in}\times \left( \cos^2(\frac{\mu_1}{2})\xi'^2 -\sin^2(\frac{\mu_1}{2})\xi''^2 \right) \frac{\partial^2}{\partial q_f'^2} \tilde \sigma_e(q_f',q_f'';t) \nonumber \\
&&\hspace{.1in}+\frac{m_e}{4\pi\hbar \delta t\cos \mu_1} \int d\xi'\int d\xi'' \nonumber \\
&&\hspace{.5in}\times \exp\left\{\frac{im_e}{2\hbar\delta t}\left( e^{i\mu_1}\xi'^2-e^{-i\mu_1}\xi''^2\right)\right\}\nonumber \\
&&\hspace{.5in}\times \left( \cos^2(\frac{\mu_1}{2})\xi''^2 -\sin^2(\frac{\mu_1}{2})\xi'^2 \right) \frac{\partial^2}{\partial q_f''^2} \tilde \sigma_e(q_f',q_f'';t) \nonumber \\
&&\hspace{.1in}+\frac{im_e\sin \mu_1}{4\pi\hbar \delta t\cos \mu_1} \int d\xi'\int d\xi'' \nonumber \\
&&\hspace{.5in}\times \exp\left\{\frac{im_e}{2\hbar\delta t} \left( e^{i\mu_1}\xi'^2-e^{-i\mu_1}\xi''^2\right)\right\}\nonumber \\
&&\hspace{.5in}\times \left(\xi''^2 -\xi'^2 \right) \frac{\partial^2}{\partial q_f'\partial q_f''} \tilde \sigma_e(q_f',q_f'';t) \ ,
\een
where the terms linear in $\xi'$ and $\xi''$ and the term containing mixed quadratic term $\xi'\xi''$ were dropped because they become zero after the integration.

Performing Gaussian integrations over $\xi'$ and $\xi''$ and expanding all terms up to the first order of $\delta t$, we obtain
\ben
&&\tilde \sigma_e(q_f',q_f'';t+\delta t) \approx  \Big \{ 1-\frac{i\delta t}{\hbar} \Big (U_e(\bar q',t)-U_e(\bar q'',t)\Big ) \nonumber \\
&&\hspace{.2in}-\frac{\delta t}{\hbar}\alpha(t)(\bar q'-\bar q'')^2 \nonumber \\
&&\hspace{.2in}- \frac{\delta t}{m_e}\tilde {\mathcal K}_I^{(1)}(t)(\bar q'-\bar q'')\Big(\frac{\partial}{\partial q_f'}-\frac{\partial}{\partial q_f''}\Big )  \nonumber \\
&&\hspace{.2in}-\frac{i\delta t}{m_e} \left (\tilde {\mathcal K}_R^{(1)}(t)-2\sin \mu_1 \tilde {\mathcal K}_I^{(1)}(t)\right)\nonumber \\
&&\hspace{.4in}\times (\bar q'-\bar q'')\Big(\frac{\partial}{\partial q_f'}+\frac{\partial}{\partial q_f''}\Big ) \nonumber \\
&&\hspace{.2in}+\frac{i\hbar\delta t}{2m_e} \left (\frac{\partial^2}{\partial q_f'^2}-\frac{\partial^2}{\partial q_f''^2}\right )\nonumber \\
&&\left . +\frac{\hbar \delta t}{2m_e} \sin \mu_1 \Big (\frac{\partial^2}{\partial q_f'^2}+\frac{\partial^2}{\partial q_f''^2}+2\frac{\partial^2}{\partial q_f' \partial q_f''} \Big )\right \} \tilde \sigma_e(q_f',q_f'';t) \ . \nonumber \\
\een
Taking the limit of $\delta t\rightarrow 0$, and making the replacement of $q_f',\bar q' \rightarrow q'$ and $q_f'',\bar q''\rightarrow q''$ in the resulting equation, we obtain the following time evolution equation:
\ben
&&\frac{\partial}{\partial t}\tilde  \sigma_e(q',q'';t)=\Big \{ \frac{i\hbar}{2m_e}\Big (\frac{\partial}{\partial q'^2} -\frac{\partial^2}{\partial q''^2}\Big )\nonumber \\
&& -\frac{i}{\hbar} \Big (U_e(q',t)-U_e(q'',t)\Big) -\frac{\alpha (t)}{\hbar}(q'-q'')^2 \nonumber \\
&&-\frac{\tilde{\mathcal K}_I^{(1)}(t)}{m_e} (q'-q'') \Big (\frac{\partial }{\partial q'} -\frac{\partial}{\partial q''}\Big )  \nonumber \\
&&-\frac{i}{m_e}\left (\tilde {\mathcal K}_R^{(1)}(t) -2\sin \mu_1 \tilde {\mathcal K}_I^{(1)}(t)\right) (q'-q'') \Big (\frac{\partial }{\partial q'} +\frac{\partial}{\partial q''}\Big ) \nonumber \\
&& +\frac{\hbar}{2m_e(t)}\sin \mu_1\Big (\frac{\partial^2}{\partial q'^2}+\frac{\partial^2}{\partial q''^2}+2\frac{\partial^2}{\partial q' \partial q''}\Big )  \Big\}\tilde \sigma_e(q',q'';t) \ . \nonumber \\ \label{eq:dt-dsigmae-1}
\een
where $\mu_1$ and $m_e$, respectively defined by Eqs. (\ref{eq:gamma1_def}) and (\ref{eq:me_def}), are all time dependent although not shown explicitly. 

\subsubsection{Case for  ${\mathcal K}_R^{(2)}(t)/m_e > 1$}

For this case, we can introduce $\mu_2(t)$ such that
\be
\sin(\mu_2(t))=\frac{m_e}{ {\mathcal K}_R^{(2)}(t)} \ . \label{eq:gam2-def}
\ee
Solving the eigenvalue problem for the quadratic form, Eq. (\ref{eq:q_exp}), it is easy to find out the following two normal modes defined in the complex domain.
\ben
&&u'=\frac{1}{\sqrt{2\cos \mu_2}} \left ( e^{i\mu_2/2} \delta q' +e^{-i\mu_2/2}\delta q''\right)\ , \\
&&u''=\frac{1}{\sqrt{2\cos \mu_2}} \left (e^{-i\mu_2/2}\delta q'-e^{i\mu_2/2} \delta q'' \right)\  .
\een
Equivalently, $\delta q'$ and $\delta q''$ can be expressed in terms of $u'$ and $u''$ as follows:
\ben
&&\delta q'=\frac{1}{\sqrt{2\cos \mu_2}} \left ( e^{i\mu_2/2} u' +e^{-i\mu_2/2} u''\right)\ , \\
&&\delta q''=\frac{1}{\sqrt{2\cos \mu_2}} \left (e^{-i\mu_2/2} u'-e^{i\mu_2/2} u'' \right)\  .
\een
Inserting these expressions into Eq. (\ref{eq:q_exp}), we find that
\be
Q(\delta q',\delta q'')=-\frac{m_e}{2\hbar\delta t}\left \{\tan (\frac{\mu_2}{2}) u'^2+\cot(\frac{\mu_2}{2})u''^2\right\} \ .
\ee
In the exponent of the integrand in Eq. (\ref{eq:rhoe-2}), squares can be completed introducing the following variables:
\ben
&&\xi'=u'+\frac{2i\delta t}{m\sqrt{2\cos\mu_2}} \cot (\frac{\mu_2}{2}) (\bar q'-\bar q'')\nonumber \\
&&\hspace{.2in}\times \left (\tilde {\mathcal K}_I^{(1)}(t)\cos(\frac{\mu_2}{2})-\tilde {\mathcal K}_R^{(1)}(t)\sin (\frac{\mu_2}{2})\right) \ , \\
&&\xi''=u''+\frac{2\delta t}{m\sqrt{2\cos\mu_2}} \tan (\frac{\mu_2}{2})(\bar q'-\bar q'') \nonumber \\
&&\hspace{.2in}\times \left (\tilde {\mathcal K}_I^{(1)}(t)\sin(\frac{\mu_2}{2})-\tilde {\mathcal K}_R^{(1)}(t)\cos (\frac{\mu_2}{2})\right) \ , 
\een
Thus, Eq. (\ref{eq:rhoe-2}) can be expressed as 
\ben 
&&\tilde \sigma_e(q_f',q_f'';t+\delta t) \approx \frac{m_e}{2\pi\hbar \delta t } \int d\xi' \int d\xi'' \tilde \sigma_e(q',q'';t)\nonumber \\
&&\hspace{.4in}\times  \exp \Big \{-\frac{m_e}{2\hbar\delta t} \Big (\tan(\frac{\mu_2}{2}) \xi'^2 +\cot (\frac{\mu_2}{2}) \xi''^2\Big )\nonumber \\
&&\hspace{.8in}-\frac{\delta t}{\hbar}\alpha (\bar q'-\bar q'')^2 \nonumber \\
&&\hspace{.8in}-\frac{i\delta t}{\hbar} \Big (U_e(\bar q',t) - U_e(\bar q'',t) \Big )  \Big \} \ , \label{eq:rhoe-s1}
\een
where 
\be 
\alpha (t) ={\mathcal K}_R^{(0)}(t)-\frac{2}{m_e}\tilde {\mathcal K}_I^{(1)}(t)\Big (\tilde K_R^{(1)}(t)-\frac{1}{\sin \mu_2} \tilde K_I^{(1)}(t)\Big)\ . \label{eq:alpha2-def}
\ee
Comparing the definitions of Eqs. (\ref{eq:gamma1_def}) and (\ref{eq:gam2-def}), it is easy to see that the above expression is equivalent to Eq. (\ref{eq:alpha1-def}). 

In Eq. (\ref{eq:rhoe-s1}), $q'$ and $q''$ can be expressed as 
\ben
&&q'=q_f'-\frac{1}{\sqrt{2\cos \mu_2}} \left ( e^{i\mu_2/2} \xi'+e^{-i\mu_2/2} \xi''\right) \nonumber \\
&&\hspace{.4in}-\frac{\delta t}{m_e} {\mathcal K}_c(t)(\bar q'-\bar q'') \ , \label{eq:r'-n1}\\
&&q''=q_f''-\frac{1}{\sqrt{2\cos \mu_2}} \left ( e^{-i\mu_2/2} \xi'-e^{i\mu_2/2} \xi''\right) \nonumber \\
&&\hspace{.4in}+\frac{\delta t}{m_e} {\mathcal K}_c^*(t)(\bar q'-\bar q'') \ , \label{eq:r''-n1}
\een 
where ${\mathcal K}_c(t)$ can be expressed as
\be
{\mathcal K}_c(t)=\tilde {\mathcal K}_I^{(1)}(t)+i\left (\tilde {\mathcal K}_R^{(1)}(t)-\frac{2}{\sin \mu_2} \tilde {\mathcal K}_I^{(1)}(t)\right) \ .
\ee
Inserting Eqs. (\ref{eq:r'-n1}) and (\ref{eq:r''-n1}) into Eq. (\ref{eq:rhoe-s1}) and following the same procedure as deriving Eq. (\ref{eq:dt-dsigmae-1}), it is straightforward to show that 
\ben
&&\frac{\partial}{\partial t}\tilde  \sigma_e(q',q'';t)=\Big\{ \frac{i\hbar}{2m_e}\Big (\frac{\partial}{\partial q'^2} -\frac{\partial^2}{\partial q''^2}\Big ) \nonumber \\
&&\hspace{.2in}-\frac{i}{\hbar} \Big (U_e(q',t)-U_e(q'',t)\Big) -\frac{\alpha (t)}{\hbar}(q'-q'')^2 \nonumber \\
&&\hspace{.2in} -\frac{\tilde{\mathcal K}_I^{(1)}(t)}{m_e} (q'-q'') \Big (\frac{\partial }{\partial q'} -\frac{\partial}{\partial q''}\Big )  \nonumber \\
&&\hspace{.2 in}-\frac{i}{m_e}\left (\tilde {\mathcal K}_R^{(1)}(t) -\frac{2}{\sin \mu_2} \tilde {\mathcal K}_I^{(1)}(t)\right) (q'-q'') \Big (\frac{\partial }{\partial q'} +\frac{\partial}{\partial q''}\Big ) \nonumber \\
&&\hspace{.2in}  +\frac{\hbar}{2m_e\sin\mu_2}\Big (\frac{\partial^2}{\partial q'^2}+\frac{\partial^2}{\partial q''^2}+2\frac{\partial^2}{\partial q' \partial q''}\Big )  \Big\}\tilde \sigma_e(q',q'';t) \ , \nonumber \\ \label{eq:dt-dsigmae-2}
\een
Comparing the definitions of Eqs. (\ref{eq:gamma1_def}) and (\ref{eq:gam2-def}), it is easy to show that the above equation is equivalent to Eq. (\ref{eq:dt-dsigmae-1}).

\end{document}